\documentclass{phbauth}      
\usepackage{graphicx}

\begin{document}

\begin{frontmatter}

\title{Vortex core size in $^3$He-$^4$He films with 
monolayer superfluid $^4$He}

\author{Han-Ching Chu},
\author{Gary A. Williams\thanksref{thank1}}
\address{Department of Physics and Astronomy, University of 
California, Los Angeles, CA 90095, USA}

\thanks[thank1]{Corresponding author. E-mail: gaw@ucla.edu} 

\begin{abstract}
The superfluid transition of 
$^3$He-$^4$He mixture films adsorbed on alumina powder is studied, 
with a $^4$He superfluid coverage near one layer.  With up to 1.3 layers 
of $^3$He added, the transition becomes strongly broadened, indicating a
linear increase in the vortex core size for $^3$He coverages below 
one layer.  Annealing of the sample mixture at 4.2 K is found 
to be critically important in ensuring a homogeneous film across the 
porous substrate.
\end{abstract}


\end{frontmatter}

Studies of the superfluid phase transition of helium films adsorbed 
in porous materials provide the only technique known for measuring 
the vortex core size in films \cite{cho,3he}.  At length scales smaller 
than the pore size of the material the film is two-dimensional, 
whereas at larger scales the film is multiply connected and becomes 
three-dimensional \cite{chan}.  The broadening of the 
transition in the 2D to 3D crossover region is essentially that of a 
finite-size Kosterlitz-Thouless (KT) transition \cite{kw,gandw,shirahama}, and 
depends on the ratio of the pore size to the vortex core size.
For a fixed pore size the degree of the transition broadening allows 
the core size to be studied as a function of film thickness and $^3$He
coverage \cite{3he}.

We report here measurments on films with a $^4$He superfluid 
coverage (at T = 0) of d$_{4}$ = 0.95 layers, higher than in our 
earlier studies using 
submonolayer coverages \cite{cho,3he}.  The same torsion oscillator 
technique and slip-cast alumina powder substrate are employed as in 
the previous measurements.  The pore size of the substrate is 
estimated \cite{beamish} to be about 100 \AA , smaller 
than the nominal powder diameter of 500 \AA \, because the strong 
surface tension forces in the slip-casting process \cite{cho}
efficiently pack the smaller powder grains into the spaces around the 
larger ones.

Figure 1 shows the measured oscillator period shift of the pure 
$^4$He film, and the same 
film for increasing coverages of $^3$He.  As discussed in Refs. 1 and 
2 these are easily calibrated in terms of the areal superfluid density, 
and as found before the transitions scale with the universal 
KT line, but are broadened above it as the 
transition crosses over to 3D.  The increasing broadening of the 
transition as $^3$He is added is readily evident in the figure. 

\begin{figure}[btp]
\begin{center}\leavevmode
\includegraphics[width=1.0\linewidth]{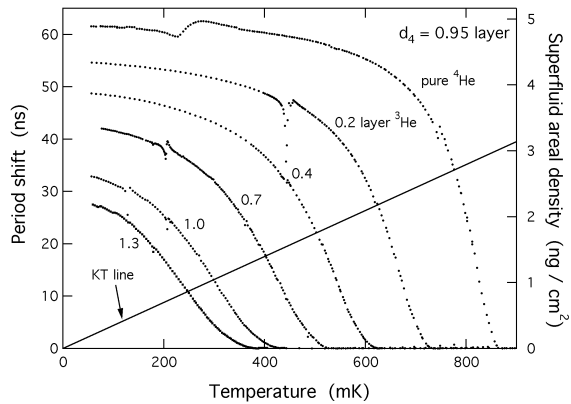}
\caption{Oscillator period shift, calibrated to give the areal 
superfluid density for the pure $^4$He film and the 
mixture films with up to 1.3 layers of $^3$He added. 
}\label{figure1}\end{center}\end{figure}

Fitting the data to the finite-size KT model used in the earlier 
work \cite{cho,3he} yields the vortex core radii shown in Fig. 2.
The present results are the solid circles, while the open circles
are the previous results \cite{3he} for a film with a thinner $^4$He
coverage, d$_{4}$ = 0.55 layers.  Both show a linear increase in the 
core size with $^3$He, but with a lower slope for the monolayer $^4$He
film.  There appears to be a leveling of the increase in the core 
size for $^3$He coverages above one layer, as might be expected since 
any further $^3$He will sit well above the $^4$He layer and only 
weakly interact with it.  This is also consistent with the reduced 
suppression of the superfluid density in Fig. 1 for the curve with 1.3
layers of $^3$He, compared to the changes in the films with 0.7 and 
1.0 layers.  Further measurements currently in progress with higher 
$^3$He coverages appear to show little further change in the core size
(to be reported elsewhere \cite{3he}).

We see no evidence in 
this data of any lateral puddling or phase separation of the $^3$He.
We also do not find any evidence of a "second transition" observed in 
other studies \cite{gasparini}.  On warming the mixture films to well
above T$_{c}$ the only observable change in the period shift is that
above about 0.8 K a slight increase in the period shift
is seen due to evaporation of $^3$He from the film into the vapor.

\begin{figure}[btp]
\begin{center}\leavevmode
\includegraphics[width=1.0\linewidth]{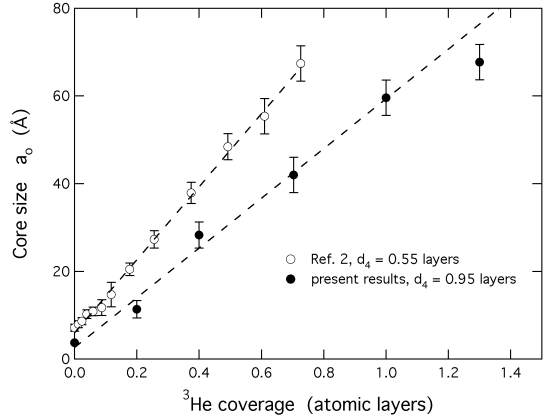}
\caption{Vortex core radius versus $^3$He coverage. 
}\label{figure2}\end{center}\end{figure}

After each increment of $^3$He was added to the $^4$He film, the cryostat was 
warmed to 4.2 K and left there for more than 24 hours to anneal the 
$^3$He and ensure that it is uniformly distributed across the 
substrate.  We have found that this is a critically important step for 
obtaining reliable data.  On one run we added 0.4 layers of 
$^3$He to the $^4$He film, and omitted the annealing step.  At higher 
temperatures the data were similar to the 0.4 layer data of Fig.1 
(shifted up in temperature by about 30 mK), 
but below 0.2 K there was a very abrupt and very large increase
in the period shift that even exceeded the period shift of the pure 
$^4$He film.  It is clear that the $^3$He was nonuniformly 
distributed in the oscillator; any
measurements made in the past with mixture films in porous materials
that did not include an annealing step probably cannot be regarded as
reliable.

This work is supported by the U. S. National Science Foundation, 
DMR 97-31523.

\end{document}